# Corrosion of Si, C, and SiC in molten salt

Jianqi Xi[1], Hao Jiang[1], Cheng Liu[1], Dane Morgan[1], Izabela Szlufarska[1,2]*

[1.]*Department of Materials Science and Engineering, University of Wisconsin, Madison, WI, 53706, USA.*
[2.]*Department of Engineering Physics, University of Wisconsin, Madison, WI, 53706, USA*

Corrosion of Si, C, and SiC in fluoride salt has been studied by *ab initio* molecular dynamics. The standard dissolution potential for Si is found to be smaller (easier to corrode) than that of C. The dissolved Si attracts $F^-$ ions and forms $SiF_6^{2-}$, whereas the dissolved C species forms neutral $CF_4$ molecules. A swapping mechanism is identified for the initial corrosion stage, where Si first comes to the surface and then is dissolved, leaving behind chain- and ring-like C structures. A strategy to suppress SiC corrosion is also discussed based on Be doping, including avoiding $Be_2C$ formation.

## 1. Introduction

Due to its excellent high temperature mechanical properties, outstanding thermo-physical/chemical properties, and low activation to neutron-induced radioactivity [1-3], silicon carbide (SiC) has a variety of potential applications in advanced nuclear reactors [2, 3]. For example, SiC has been proposed as a coating for nuclear graphite, which is used for neutron moderator and a reflector. In molten salt reactor (MSR), typically a fluoride salt, e.g., FLiBe (66 mol% LiF-34 mol% $BeF_2$), is used as a fuel solvent or coolant [4, 5]. Unlike in oxidizing environments [6], SiC exposed to fluoride salt does not form a protective oxide layer, and mechanisms and rates of degradation of SiC in fluoride salt are still not well understood. Interesting fundamental questions are the standard potentials of Si and C, dissolved Si and C speciation, how bare surfaces of SiC are dissolved due to a direct contact with corrosive salts and whether there might be a non-oxide protective layer formed on the surface in these environments.

Recently, behavior of SiC in fluoride salt was investigated by Gu *et al.* using a combination of scanning electron microscopy, X-ray diffraction, and X-ray photoelectron spectroscopy techniques. The authors reported that SiC does corrode in the salt, and the corroding surface becomes depleted in Si content. This result suggests that Si might have been dissolved into the salt, resulting in the formation of a carbon-rich surface [7]. Similar results have also been reported by Xue *et al.* [8] who found that impurities from the salt, such as $Cr^{3+}$, could further accelerate corrosion of SiC. These results suggested that unlike the oxide protected surface formed in oxidizing environment, the SiC surface is corroded in fluoride molten salt via dissolving Si species. Although these experimental observations provide some information on the corrosion behavior, the understanding of interactions between molten salt and SiC surface is still very limited. For instance, it is not known by what mechanism Si can dissolve preferentially from the multi-component SiC, whether the carbon-rich layer is always expected to form in fluoride salt, how stable is this layer under different salt conditions, and finally it remains to be determined whether the chemistry of salt can be tuned to suppress SiC corrosion?

---
* *Department of Materials Science and Engineering, University of Wisconsin, Madison, WI, 53706, USA.*
E-mail: szlufarska@wisc.edu; *Tel: +1-608-265-5878*

To address the above questions, here we use *ab initio* molecular dynamics (AIMD) simulations within the framework of density functional theory (DFT). We focus on the FLiBe salt, which is considered to be a primary coolant or fuel salt in MSR. A comparison to behavior of SiC in another commonly used salt, FLiNaK, is also included in the discussion part of this paper. We choose 3C-SiC(001) as a representative surface to investigate the dissolution mechanism of Si during corrosion. We consider surfaces with both Si- and C-termination and with $p(2\times1)$ surface reconstructions, which are known to be stable [9, 10]. Our model surfaces are shown in Fig. 1. The following issues are considered in this work: (i) we calculate the standard potential for Si, C, and SiC dissolution in fluoride molten salt and we determine the configuration of dissolved species in the corrosion; (ii) We identify steps in the kinetic pathway for Si dissolution; (iii) Based on these results, we discuss possible strategies for controlling salt activity in SiC.

One should note that while standard potentials for Si and C solids are useful to calculate in order to understand corrosion of SiC, corrosion behaviors of Si and C in molten salts are interesting in their own right for fields such as chemical processing technology and device applications [11-16]. For instance, processing of Si by molten salts has been suggested to be more cost efficient than conventional processing by the carbothermic reduction of $SiO_2$ [11-14]. In another example, the electrochemical exfoliation of cathodically charged graphite materials in molten salt has been proposed as an efficient approach for economic and environmentally sustainable production of high quality graphene-derived nanostructures [15, 16]. Therefore, although here our primary interest is on corrosion of SiC, we also present thermodynamic calculations for silicon and graphite solids.

## 2. Methodology

## 2.1 Computational details

To prepare a model of a molten salt, we first create a cell containing 98 atoms with a eutectic composition of $BeF_2$-2LiF (56 F atoms, 28 Li atoms, and 14 Be atoms). Positions of atoms are random, and they are chosen using Packmol code [17]. Before doing AIMD simulations, the ionic positions are pre-equilibrated using classical molecular dynamics (MD) with Born-Meyer-Higgins type potentials [18-20] using Large-scale Atomic/Molecular Massively Parallel Simulator (LAMMPS) code [21]. The final structure from classical MD simulations is used as an input for AIMD simulations. The classical MD simulations are carried out as follows [18]. We use the time step of 1 fs for all the simulations. We first run 5 ps of simulations in the canonical ensemble (NVT). The temperature here is controlled using the Langevin thermostat, which quickly brings the system to the desired temperature. Next, we use an isothermal-isobaric (NPT) ensemble with Nose-Hoover thermostat and barostat for 100 ps, which equilibrates the system to the desired pressure and temperature. From these simulations, we obtain an equilibrium volume by averaging lattice parameters over the last 25 ps of simulation. Using this volume, we then perform simulations in NVT ensemble with the Nose-Hoover thermostat, running for another 100 ps. The final structure is used to initialize the AIMD simulations. In the AIMD simulations, we use the NVT ensemble with the Nose-Hoover thermostat with a time step of 1 fs. Simulations are performed using the Vienna Ab Initio Simulation Package (VASP) [22]. Settings of the DFT simulations will be discussed below. The structure and density for the fully relaxed system after 20 ps at 1000 K are validated against previously published results, as summarized in Table 1.

Table 1. The comparison of the distance for ion pairs and density of FLiBe with the previous results. Present work for FLiBe are performed at 1000 K.

|  | Present work | Previous AIMD[a] | Exp |
|---|---|---|---|
| Be-F (Å) | 1.55 | 1.55 | 1.58[b] |
| Li-F (Å) | 1.87 | 1.88 | 1.85[b] |
| F-F (Å) | 2.62 | 2.58 | 2.56-3.02[b] |
| Density | 1.99 | 1.97 | 1.93[c] |
| (g/cm$^3$) |  |  | 2.08[d] |

[a] AIMD simulations at 973 K, Ref [19]
[b] Analyzed from X-ray diffraction data measured at 1123 K, Ref [23]
[c] Ref [24]
[d] Ref [25]

Projector-augmented-wave (PAW) potentials [26] are used to mimic the ionic cores, while the generalized gradient approximation (GGA) in the Perdew-Burke-Ernzerhof (PBE) [27] approach is employed for the exchange and correlation functional. In the calculations, we test the dispersion effect with vdW-optB88 term [28], as discussed in Fig. S1. The optimized coordinates and lattice parameter for SiC are taken from our earlier studies [29].

The SiC(001) surfaces are represented by a symmetric slab with 13 layers and 8 atoms per layer along the (001) direction, with two identical free surfaces at both ends. Periodic boundary conditions are employed in all three spatial directions. The vacuum region with ~15.1 Å is filled with the pre-relaxed molten salt atoms. Previous calculations have confirmed that the size effects for SiC(001) surfaces can be negligible when the slab is larger than 11 layers with a 12.5 Å vacuum region [15].

All computations are performed with a cutoff energy of 500 eV for the plane wave basis set and with spin-polarized conditions. The integration over the Brillouin zone is performed using the Γ point for AIMD simulations, and the 3×3×1 kpoint mesh is used for the subsequent energy minimization. Motivated by experimental work at high temperatures (around 1000 K) [7, 8], the AIMD simulations are performed at 1000 K for the thermodynamics calculations (see Sec. 2.2). Kinetics of dissolution is simulated at 1000 K and 2000 K, as the higher temperature is expected to accelerate the reaction so that it could be observed within the time scale of AIMD.

## 2.2 Standard dissolution potential in fluoride salt

To model the electrochemical dissolution reactions in the fluoride salt we build on the approach described by Nam, *et al.* [19] and extend it to include references to pure elements. In fluoride salt, the electrochemical reaction involved in the dissolution of species X (where X stands for Si or C) in SiC can be written as

$$X^{SiC} + salt \leftrightarrow X^{n+}_{salt} + ne^-$$ (1)

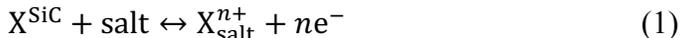

Here, $X_{salt}^{n+}$ is the dissolved species in the salt and $X^{SiC}$ is the species in SiC, $n$ is the number of electrons involved in the reaction. The energy of an electron in equilibrium with the above reaction is referred to as the electrochemical potential and for a specific case, where $X_{salt}^{n+}$ is at 1 molar concentration, this energy is called the standard potential, $E^0$. Generally, the more negative the standard potential, the easier it is to dissolve the species and to corrode the material. The reference electrode (i.e., the chemical potential) for electrons in Eq. (1) is chosen to be the $F_2|F^-$ redox reaction in the FLiBe salt at 1 atm partial pressure of $F_2$, which can be written as

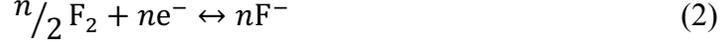

$$^n/_2 F_2 + ne^- \leftrightarrow nF^- \qquad (2)$$

The half-cell reactions given by Eqs. (1) and (2) can be combined to give

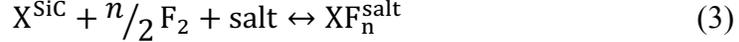

$$X^{SiC} + ^n/_2 F_2 + salt \leftrightarrow XF_n^{salt} \qquad (3)$$

Since there are no electrons gained or lost in this reaction, it is possible to determine its free energy based on DFT calculations. Specifically, the reaction Gibbs free energy of reaction in Eq. (3) can be written as

$$\Delta G = \langle E(XF_n^{salt}) \rangle - \langle E(salt) \rangle - \mu(X^{SiC}) - ^n/_2 \mu(F_2) + \Delta C_s \qquad (4)$$

where $\langle E(XF_n^{salt}) \rangle$ is the ensemble average of the total internal energy of the product, $XF_n$, dissolved in the salt. $\langle E(salt) \rangle$ is the ensemble average of the total internal energy of pure salt. $\mu(X^{SiC})$ is the chemical potential of Si or C species in SiC at the temperature where the potential is being measured, $\mu(F_2)$ is the chemical potential for fluorine gas at the temperature of the calculation and 1 atm pressure. $\Delta C_s$ is the entropy correction term, describing the change of entropy of the product, $XF_n$, from its standard state (e.g., for $SiF_4$ the standard state would be that gas phase at standard temperature and pressure) to the dissolved one. $\Delta C_s$ is taken to be the same as the change in entropy of the standard state to the stable phase of the corrosion products at the temperature of the simulation and one atmosphere (e.g., for simulation of $SiF_4$ at 1000K this would be the gas phase of $SiF_4$ at 1000K and one atmosphere) [30]. This approach is reasonable if we assume that the entropy of the species is not altered significantly by being moved from the gas phase to the salt. Once $\Delta G$ is calculated from Eq. (4), we can relate it to the standard dissolution potential as $E^0 = \Delta G/_{nF}$, where $F$ is the Faraday constant.

To obtain the ensemble average of the total internal energy, we performed AIMD simulations in the NVT ensemble for 20 ps in each case with 1 fs time step at the temperature of 1000 K. Averages are calculated during simulations that follow a 5 ps equilibration period. For the modeling of the dissolved product in salt, the dissolved cation with excess anions are simultaneously inserted in the simulation cell. For example, for the dissolved product of $SiF_4$, one $Si^{4+}$ and four $F^-$ are added to the 98 atoms of the FLiBe salt cell, so that the cell maintains its neutrality. The effects of system size have been tested by using a larger supercell with 395 atoms. We found that the potential difference between large and small supercell is around 0.1 V, which is negligible on the scale of trends reported in this paper. The above method for calculating energies of dissolved species has been successfully applied before to calculate the solute behavior in molten salts [30]. We have also tested the effect of dispersion correction on the calculated standard potential, and we found this effect to be negligible, as shown in Fig. S1. The chemical potential of Si and C, $\mu(Si^{SiC})$ and $\mu(C^{SiC})$ in Eq. (4) are related to the chemical potential of SiC compound and constrained by $\mu(SiC^{SiC}) = \mu(Si^{SiC}) + \mu(C^{SiC})$. In this work, the Si- and C-rich limits for Si and C correspond to the bulk silicon and graphite, respectively. The chemical potentials for Si and C are assumed to be constrained to be between the Si-rich limit and the C-rich limit, which are shown in Table 2. The temperature effect for chemical potential of the solid is included as follows:

$$\mu_{solid}(T) = \mu_{solid}(0\ K) + \big(H(T) - H(0\ K)\big) - T\big(S(T) - S(0\ K)\big) \qquad (5)$$

$\mu_{solid}(0\ K)$ is the reference state of solid from our published DFT calculations [29, 31], where zero for C and Si are set by the VASP pseudo-atom energy. $H$ and $S$ are the enthalpy and entropy, respectively, which have been experimentally determined and tabulated in JANAF database [32]. The chemical potential of fluorine gas is determined by referencing the JANAF database using the approaches from Refs. [33, 34] via putting one $F_2$ gas molecule in a vacuum box, and the values for $\mu(F_2)$ at 0 K and 1000 K and standard pressure are -3.52 and -5.70 eV, respectively.

*Table 2. The chemical potential of solid Si and C species in SiC as well as Be metal at 0 K and 1000 K. The unit is eV/atom.*

|  | $\mu(Si^{SiC})$ | | $\mu(C^{SiC})$ | | $\mu(Be^{metal})$ | |
|---|---|---|---|---|---|---|
|  | 0 K | 1000 K | 0 K | 1000 K | 0 K | 1000 K |
| Si-rich | -5.43 | -5.71 | -9.63 | -9.96 | -3.75 | -3.93 |
| C-rich | -5.73 | -6.22 | -9.33 | -9.45 | | |

## 2.3 Concentration of metallic Be addition in fluoride salt

The thermodynamics of Be addition will play an important role in this work as it can buffer the salt and potentially interact with SiC. Specifically, we will need a model for dissolution of Be metal as neutral Be ($Be^0$) in the salt. To build this model we note that the change in Gibbs free energy for neutral $Be^0$ dissolved in the salt phase, $Be^0_{salt}$, can be calculated as

$$\mu\big(Be^0_{salt}\big) - \mu\big(Be^{metal}\big) = \langle E\big(Be^0_{salt}\big)\rangle - \langle E(salt)\rangle - \mu\big(Be^{metal}\big) - T\Delta S \qquad (6)$$

where $\mu\big(Be^0_{salt}\big)$ and $\mu\big(Be^{metal}\big)$ are the chemical potential of $Be^0$ species in the salt and metallic phase, respectively. $\langle E\big(Be^0_{salt}\big)\rangle$ is the ensemble average of the total internal energy of dissolved $Be^0$ species in the salt, which can be obtained in our AIMD simulations as discussed in Sec. 2.2. $\Delta S$ is the entropy for the dissolution of $Be^0$ species into the salt. Based on the Raoult's law, it is estimated as

$$\Delta S = -k_B ln\big(C^{salt}_{Be^0}\big) \qquad (7)$$

where $k_B$ is the Boltzmann constant, and $C^{salt}_{Be^0}$ is the mole fraction of dissolved $Be^0$ species in the salt. A parallel formalism to Eq. (5) with a DFT calculations of Be is used for the evaluation of $\mu\big(Be^{metal}\big)$, which is given in Table 2. Combining Eqs. (6) and (7), the concentration of dissolved $Be^0$ species in fluoride salt can be determined by knowing the value of $\mu\big(Be^0_{salt}\big)$. For instance, the maximum solubility limit of $Be^0$ species in the salt when in contact with Be metal at 1000 K can be estimated as $\mu\big(Be^0_{salt}\big) = \mu\big(Be^{metal}\big) = -3.93$ eV, which is the highest possible chemical potential for $Be^0$ species in the salt. In this case, the concentration of neutral Be species dissolved in the salt is ~$2.51\times10^{-12}$ mol% (where mol% $x = {n_x}/{(n_{F^-} + n_{Li^+} + n_{Be^{2+}} + n_x)} \times 100\%$, and $n_i$ is number of moles of species $i$).

We will also be interested in the reaction of Be with C in the SiC to form solid $Be_2C$ through the reaction

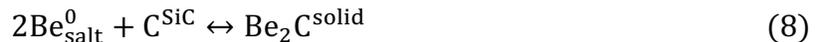

$$2Be^0_{salt} + C^{SiC} \leftrightarrow Be_2C^{solid} \qquad (8)$$

The Gibbs free energy of reaction in Eq. (8) can be written as

$$\Delta G_f = \mu(\text{Be}_2\text{C}^{\text{solid}}) - 2\mu(\text{Be}^0_{\text{salt}}) - \mu(\text{C}^{\text{SiC}}) \qquad (9)$$

where $\mu(\text{Be}_2\text{C}^{\text{solid}})$ is the chemical potential of solid $\text{Be}_2\text{C}$ in the $Fm\overline{3}m$ phase [43]. Again, a parallel formalism to Eq. (5) with a DFT calculations of $\text{Be}_2\text{C}^{\text{solid}}$ is used for the evaluation $\mu(\text{Be}_2\text{C}^{\text{solid}})$, which yields -17.89 eV/(formula unit) for the energy relative to graphite and diamond phase Si.

# 3. Results

## 3.1 Thermodynamics of Si, C, and, SiC corrosion

The standard potential for solid dissolution in salt is an important thermodynamic factor that determines how the material is dissolved in salt. Here, we calculate the standard potential for dissolution of silicon, graphite, and SiC in fluoride salt, and we compare the results with the experimentally measured salt potential. The results are shown in Fig. 2. In Table S1 of Supplementary we have also compared the calculated results with the thermodynamic database [32]. From the data in Fig. 2(a), it is clear that the standard potential for silicon dissolution in fluoride salt is smaller (more negative) than that of graphite, which suggests that graphite is more stable than silicon in fluoride salt. These results are consistent with thermodynamic database [32], which shows that $F_2$ reacts more strongly with silicon than with graphite. For SiC dissolution (see Fig. 2(b)), we find similar results, i.e., the standard potential for Si species (yellow region) dissolved from SiC into the salt is smaller than that of C species (blue region), regardless of whether we consider Si- or C-rich reference states. Thus, our results are consistent with the hypothesis that Si dissolves more easily than C, and SiC might be able to build up a C-rich layer on the surface.

In addition to the calculations of standard potentials, we have used AIMD simulations to identify configurations of species dissolved in the salt (details of calculations can be found in Sec. 2.2). For Si species, we found that the dissolved $Si^{4+}$ cation interacts six $F^-$ anions and forms $SiF_6^{2-}$, in which six anions form an octahedron around the central $Si^{4+}$ cation with $O_h$ symmetry, and the Si-F bond length is around 1.70 Å (see Figs. 3(a) and (c)). Similar results have been experimentally reported for SiC in FLiNaK salt where the $Na_2SiF_6$ and $K_2SiF_6$ were created due to corrosion [8]. Apart from the most energetically favorable reaction of $Si|Si^{4+}$, according to Fig. 2, there are other dissolution reactions that could potentially take place. Specifically, $Si|Si^{2+}$ and $Si|Si^{3+}$ have standard potential within the range of experimental values and therefore might take place during the corrosion of SiC in salt. Using AIMD, we found that configurations of dissolved species for the latter two reactions are $SiF_3^-$ (with $C_{3v}$ symmetry) and $SiF_4^-$ (with $C_{2v}$ symmetry) in FLiBe salt, respectively, with bond lengths around 1.66 - 1.73 Å (see supplementary Fig. S2). These results suggest that as Si species is dissolved into fluoride salt, it prefers to attract extra $F^-$ ions and form anion clusters. For C species, however, the dissolved $C^{4+}$ cation interacts with four $F^-$ anions, resulting in the formation of neutral $CF_4$ molecule. Specifically, one $C^{4+}$ cation occupies the center of a tetrahedral structure, and four $F^-$ ions are located at each vertex of the tetrahedron (with $T_d$ symmetry). The C-F bond length in this case is around 1.33 Å (see Figs. 3(b) and (d)). However, the high standard potential for $C|C^{4+}$ reaction relative to the measured salt potential suggests that the concentration of C species dissolved in the salt will be relatively low.

The calculated values of the standard potential for dissolution of SiC can be used to consider how to control the salt potential in applications in order to suppress the dissolution. As shown in Fig. 2, the lower limit of the experimentally measured salt potentials is smaller than the potential for Si species, suggesting that it might be possible to pin the salt potential to suppress Si dissolution. Reduction of the standard potential of fluoride salt can be achieved by adding beryllium metals into FLiBe molten salt [36]. The lowest limit of the salt potential that can be reached via this route corresponds to the potential for $Be^{metal}|Be^{2+}$, which is shown in Fig. 2 as a dashed line, around -4.61 V with standard deviation of 0.28 V. This potential has been obtained based on the approach discussed in Sec. 2.2. In this case, the standard dissolution potential for Si species would be around 1 V higher than the salt potential, which means that the dissolution of Si species would be definitely suppressed. These results suggest that theoretically there is a room in the range of salt potentials where salt chemistry can be adjusted to suppress corrosion of SiC.

## 3.2 Corrosion mechanism of SiC in Molten Salt

In the above thermodynamic calculations, we found that corrosion of SiC in fluoride salts dissolution proceeds by dissolution of Si species. However, it remains to be determined by what mechanism Si can preferentially dissolve from a binary covalent material such as SiC and whether this mechanism is kinetically accessible to typical experiments. Here, we analyze corrosion mechanisms of SiC using AIMD simulations for both C- and Si-terminated SiC(001) surfaces. The (001) surface is one of the most energetically stable surfaces in SiC [37].

### 3.2.1 C-terminated surface
*Surface electronic structure*

In order to better understand the salt-surface interaction, the electronic structure for C-terminated surface is investigated. The partial charge density of surface states with energies 1.5 eV below the Fermi level is shown in Fig. 4(a). Two criteria have been considered when determining the value of the cut-off energy. Firstly, in order to visualize the active valence electron near the surface, the cut-off energy should be around the Fermi level. Previous DFT calculations reported that the occupied surface states are mainly within the range of 2 eV below the Fermi level [38]. Secondly, the cut-off should be chosen so that one can clearly see the occupied surface states without too much subsurface noise. We have tested the energies from 0.5 eV to 2 eV below the Fermi level, and we found that the cut-off of 1.5 eV was large enough to describe the partial charge density of surface states. From Fig. 4(a) it can be seen that the occupied surface states mainly originate from the interaction of the surface layer C atoms and the second layer Si atoms. The carbon dangling bonds of C=C dimer surface is almost empty, indicating that there is no active valence electron near the surface, which is consistent with previous DFT calculations on the pure C-terminated surface [38]. These results suggest that it is difficult for anions in the salt to obtain electrons from the C=C dimer surface, and thus F atoms can rarely form the C-F bond near the surface. The following analysis of bonding states investigations further confirms these results.

In order to study bonding states of atoms at the salt/SiC interface, we have analyzed the difference in charge densities between the salt/SiC system and isolated salt and SiC surface. In the latter two cases, we have frozen the ionic structures and only relaxed the electronic structures. From Fig. 4(b), we can see that instead of interacting with the surface carbon atoms, F atoms preferentially interact with the second layer Si atoms, forming the ionic bonds with 1.84 Å bond length, along <111> direction. Based on the Ref. [39], the strength of these bonds can be measured via the integration of projected crystal orbital Hamilton population (p$COHP$). The integrated p$COHP$ ($ICOHP$) is taken here to be a measure of bond strength and will be referred to as "bond strength", and it can be calculated as follows [39]:

$$ICOHP = \int_{-\infty}^{E_F} \text{p}COHP \, \text{d}E \qquad (10)$$

Based on this analysis, we find that the bond strength of Si-F along the <111> direction of SiC (-2.52 eV) is stronger than that of Si-C bond along the same direction (-2.04 eV), which results in the shift of a Si atom out of its lattice site. In addition, the strong interaction between Si and F atoms suggests that F atoms in the salt tend to be attracted to Si in the sub-surface layer.

***Structural evolution***

Structural evolution of the atoms during the interactions of fluoride salt and C-terminated SiC(001) has been studied by running AIMD at 1000 K and 2000 K for 10 ps. The high temperature is expected to accelerate the reaction within the AIMD time scale. We find that multiple F atoms are adsorbed near the surface and forms bonds with Si atoms in the subsurface layer. The strong interactions between F and Si elongate the Si-C bond, especially along the <111> direction, from 1.85 Å to 2.15 Å, and in some cases even break these Si-C bonds. Meanwhile, the bond lengths for Si-F decrease from 1.84 Å to 1.70 Å. Finally, some Si atoms (red color in Fig. 5) are pulled out of their original lattice sites by the F atoms to which they are bonded (see Fig. 5(b)). At the same time, C atoms in the top layer (black color in Fig. 5) tend to swap positions with the second layer Si atoms, they interact with the third layer C atoms, resulting in the formation of triangular C ring structures. The distances between C atoms are 1.51 - 1.61 Å, featuring C-C single bonds, which are different from the initial C=C dimer (1.36 - 1.40 Å). The swapping reaction of Si and C atoms continues to take place, i.e., the second layer Si atoms are pulled out to the top of the surface, while the top layer C atoms move down and eventually forms the C$_{Si}$ antisites. This swapping reaction exposes subsurface Si atoms to the salt environment, which may be the prerequisite condition for Si dissolution. However, due to the limitation of the AIMD time scales, the Si dissolution is not observed in the 10 ps simulations.

In order to be able to observe in AIMD simulations how swapping mechanism leads to dissolution, we have created a more corrosive environment by adding additional F atoms into the salt, as shown in Figs. 5(c) and (d). These F are added near the middle of the salt region above the SiC slab and placed in such a way as to avoid overlap with other atoms. We found that adding four F atoms in the salt leads to more swapping reactions on the surface, and formation of a chain-like carbon structure (see Fig. 5(c)). Adding another four F atoms to the simulation cell (the total of 8 extra F) leads to dissolution of the swapped Si atom into the salt. The dissolved Si atom forms the SiF$_6^{2-}$ species, and a ring-like carbon structure is created on the surface (see Fig. 5(d)). This observation is consistent with our thermodynamic calculations and discussion in Sec. 3.1, where the most energetically favorable product of the Si|Si$^{4+}$ reaction was found to be the silicon hexafluoride structure. The chain-like and ring-like carbon structures left behind on the surface have bond length of 1.37 - 1.42 Å, and form double bonds, with $sp^2$ hybridization. These complex carbon structures could be the precursors for the growth of amorphous carbon-rich layers (with high concentration of $sp^2$ hybridization) as reported in the previous experiments [7, 8]. These results further confirm that the swapping mechanism on C-terminated surface is a necessary step for Si dissolution in fluoride salt and plays an important role in SiC corrosion.

### 3.2.2 Si-terminated surface
*Fluorination of surface*

Bonding on the Si-terminated SiC(001) surface is qualitatively different from that on the previously discussed C-terminated surface. As shown in Fig. 6(a), dangling bonds on Si on the Si-terminated surface are localized with valence electrons, which is consistent with previous DFT calculations on the pure Si-terminated surface [38]. The unpaired valence electrons on Si form electrically active interface traps, which can easily attract F atoms from the salt. As a result, ionic Si-F bonds are formed with the lengths of 1.69 - 1.71 Å (see Fig. 6(b)). Similar results have been found in our AIMD simulations at the temperatures of 1000 K and 2000 K. We find that F atoms do not perturb the structure of Si-terminated surface and simply passivate the Si dangling bonds. The same result was observed even if additional F atoms were introduced into the salt (we tested up to 12 additional F atoms). This behavior is consistent with the negligible evolution of the effective charge for the top Si layer and the second C layer, as shown in Fig. 7, suggesting that the F-passivated Si-SiC(001) surface is stable within our AIMD simulation time scale. Similar results have been observed in $SrTiO_3$ [40], Si [41], and $ZrO_2$ [42] by using fluorine treatment to passivate the active charge trapping sites of the surface.

However, due to the high electronegativity of F atom, it is likely that at times longer than our AIMD time scale, additional F atoms near the surface might cleave the Si-C bonds and lead to degradation of the Si-terminated surface. A similar trend has been reported based on experiments in $ZrO_2$ [42]. Specifically, Huang *et al.* [42] observed that $ZrO_2$ thin film was passivated by F atoms at short treatment time, whereas longer exposure time led to fluoride corrosion of the same samples. It is also important to point out that the surface in our simulations is ideal in the sense that it does not have any defects. It is possible that such surface is in fact stable against corrosion even for relatively long exposure times and that in experiments initiation of corrosion would occur at surface defects, such as vacancies, antisites, and surface steps.

## 4. Discussion and Conclusions

According to our thermodynamic calculations, corrosion of SiC in fluoride salt occurs first through dissolution of Si species with increasing potential. The calculated value of standard potential for Si dissolution allows us to consider how low a salt potential might be needed to thermodynamically suppress corrosion of SiC. One common strategy for reducing the salt potential is to add Be metal to the salt, which as shown in Sec. 3.1, may stabilize the Si species in SiC. However, it is important to consider if the introduction of beryllium species might simultaneously induce the production of solid metal carbides, of which the most stable is assumed to be $Be_2C$ [36, 43], which would be formed via Eq. (8). At the maximum solubility limit for $Be^0$ in the salt it is in equilibrium with the Be metal and the chemical potential $\mu(Be_{salt}^0) = -3.93$ eV (see Sec. 2.3). At this $\mu(Be_{salt}^0)$ the salt potential is set by the reaction $Be^{metal}|Be^{2+}$ to be around -4.61 V via the approach in Sec. 2.2, as shown the dashed line in Fig. 2. Meanwhile, the calculated $\Delta G_f$ for the formation of solid $Be_2C$ in Eq. (8) is about -0.58 eV (-0.07 eV) for the C-rich (Si-rich) condition, implying that the reaction is exothermic, and thus predicting that in equilibrium with Be metal the C species in SiC will be reactive, and form solid

Be$_2$C. A bounding line for C stability vs. Be$_2$C$^{solid}$ formation is $\Delta G_f = 0$ in Eq. (9), i.e., $\mu\left(Be_{salt}^0\right) = 1/2 \left(\mu(Be_2C^{solid}) - \mu(C^{SiC})\right)$. This equality occurs for $\mu\left(Be_{salt}^0\right)$ = -4.22 eV and -3.97 eV for the C-rich and Si-rich condition, respectively. The corresponding concentration of dissolved Be metal in the salt is ~ 8.96×10$^{-14}$ and 1.73×10$^{-12}$ mol% for the C-rich and Si-rich condition, respectively. Based on the Nernst equation [44], the salt potential in these cases, assuming it is set by the Be$_{salt}^0$|Be$^{2+}$ reaction, is shifted to -4.32 V and -4.58 V in the C-rich and Si-rich condition, respectively. These values are still lower than the standard potential for Si and C species dissolution in the salt, as shown in the green region in Fig. 2. The key aforementioned values of $\mu\left(Be_{salt}^0\right)$, $C_{Be^0}^{salt}$, salt potential, $E$, and the Gibbs free energy for the formation of solid Be$_2$C, $\Delta G_f$, in the C- and Si-rich conditions can be found in Table 3.

The above results suggest that it is at least in theory possible to reduce the salt potential with Be additions to suppress the corrosion of SiC in fluoride salt without precipitating out metal carbides by controlling the concentration of dissolved Be$^0$ species in the salt to be lower than ~ 10$^{-14}$ mol% for C-rich condition, and ~ 10$^{-12}$ mol% for Si-rich condition. Note that in the above arguments we have assumed that controlling the salt potential to below the standard potential of Si species would prevent Si dissolution. However, if the concentration of dissolved Si species, for example Si$^{4+}$ species, in the salt is lower than 1 molar, the potentials of Si dissolution would be lower than the standard potential. If we somewhat arbitrarily set the tolerable concentration for the dissolved Si$^{4+}$ species in the salt in the order of 1×10$^{-10}$ molar, then the potential to stabilize that concentration is around -4.05 V on the basis of the Nernst equation [44]. This value is significantly lower than the standard potential, but still above the lower limits set for potential control by solid Be$_2$C formation, as shown in Table 3.

*Table 3. Concentration of dissolved Be$^0$ atom in the FLiBe salt at different conditions and T = 1000K. Case 1 is for the condition of $\mu\left(Be_{salt}^0\right) = \mu(Be^{metal})$, and Case 2 is for the $\mu\left(Be_{salt}^0\right) = 1/2 \left(\mu(Be_2C^{solid}) - \mu(C^{SiC})\right)$.*

| Case | | $\mu\left(Be_{salt}^0\right)$ (eV) | $C_{Be^0}^{salt}$(mol%) | $E$ (V) | $\Delta G_f$ (eV) |
|---|---|---|---|---|---|
| 1 | C-rich | -3.93 | 2.51×10$^{-12}$ | -4.61 | -0.58 |
| | Si-rich | -3.93 | 2.51×10$^{-12}$ | -4.61 | -0.07 |
| 2 | C-rich | -4.22 | 8.96×10$^{-14}$ | -4.32 | 0 |
| | Si-rich | -3.97 | 1.73×10$^{-12}$ | -4.58 | 0 |

We also note that if solid Be$_2$C is highly soluble in the salt then potentially a significant amount of reaction of C and Be could occur without precipitating Be$_2$C$^{solid}$ but instead forming Be$_2$C$^{salt}$ through the reaction

$$2Be^{0,salt} + C^{SiC} + salt \leftrightarrow Be_2C^{salt} \tag{11}$$

However, following the approach in Sec. 2.2, the Gibbs free energy for this reaction is about 0.84 eV at 1000 K for even the most active Be condition $(\mu\left(Be_{salt}^0\right) = \mu(Be^{metal}))$ and C condition (C-rich). This high positive reaction energy suggests that this reaction will allow only a small amount of Be$_2$C$^{salt}$, with a predicted equilibrium concentration under these conditions of $C_{Be_2C}^{salt}$ is ~ 5.8×10$^{-3}$ mol%. This value of $C_{Be_2C}^{salt}$ might be expected to correlate with slow dissolution kinetics, but it is large enough that it could correspond to significant C loss. However, this value is a worst-case prediction that would be greatly reduced when consider C pulled from Si-rich SiC and could be reduced by reducing Be content.

Finally, for comparison, we have also calculated the standard potentials for SiC dissolution in another commonly used fluoride salt, FLiNaK (46 mol% LiF, 42 mol% KF, 12 mol% NaF). Calculations were carried out using the same procedure as for FLiBe. As shown in Fig. 8, the standard potentials for Si and C species in both fluoride salts are qualitatively comparable, i.e., in both salts Si species are more likely to be corroded as compared with C species. For Si species, the reaction for $Si|Si^{4+}$ is the most energetically favorable, which is followed by $Si|Si^{2+}$ reaction. For C species, the most favorable reaction is the $C|C^{4+}$ reaction. Quantitatively, however, the standard potentials in FLiNaK salt are slightly lower (more negative) than those in FLiBe salt, indicating that compared with FLiBe salt, species in SiC are more likely to lose electrons and thus easier to dissolve into FLiNaK salt. The different corrosivity of these fluoride salts has been previously reported for other materials, e.g., it was found that the corrosion rate of ferritic steel ($Fe_9Cr_2W$) in FLiNaK is ten times greater than in FLiBe [45]. It is possible that the different corrosion behavior of materials in the two salts is due to the ionic states of these salts. Our simulations show that the ionic states of the same elements can be different in the two fluoride salts. Specifically, in FLiBe salt, a network of $BeF_4^{2-}$ tetrahedral entities, such as $BeF_4^{2-}$, $Be_2F_7^{3-}$, and $Be_3F_{10}^{4-}$, are connected at their corners, with few free $F^-$ ions. In contrast, FLiNaK salt appears to have an excess of free $F^-$ ions. Based on Lewis acid-base arguments [46, 47], these different ionic states could result in FLiNaK being inherently more corrosive to solid materials than FLiBe.

In summary, the corrosion behavior of SiC in fluoride molten salts has been investigated by *ab initio* molecular dynamics simulations based on density functional theory. The standard potentials for the dissolution of SiC in fluoride salt have been determined. It is found that the potential of Si species is smaller than that of C species, suggesting that the Si species would be much easier to dissolve into the salt than the C species. We have also identified the configurations of the dissolved Si species and found the following: (i) the dissolved $Si^{4+}$ prefers to form $SiF_6^{2-}$ with $O_h$ symmetry, in which six anions surround the central $Si^{4+}$ cation in octahedral configuration, and the Si-F bond length is around 1.70 Å; (ii) the dissolved $Si^{2+}$ and $Si^{3+}$ form $SiF_3^-$ ($C_{3v}$ symmetry), and $SiF_4^-$ ($C_{2v}$ symmetry), respectively, with bond lengths around 1.66 - 1.73 Å. These results suggest that as Si species dissolves into fluoride salt, it prefers to attract additional $F^-$ ions and to form anion clusters. For C species, owing to the high standard potential, only few $C^{4+}$ species may dissolve into the salt. When dissolved C atom forms a neutral $CF_4$ molecule with $T_d$ symmetry i.e., one $C^{4+}$ cation occupies the center of the tetrahedral structure, and four $F^-$ ions are located at each vertex. The C-F bond lengths for these configurations are around 1.33 Å.

We have also elucidated the mechanisms of SiC corrosion in fluoride salt. For a C-terminated surface, we found a swapping mechanism that brings subsurface Si atoms to the surface, which is a necessary step for Si dissolution and for formation of C-rich layer during corrosion. For Si-terminated surface, we found that anions form a passivation layer and no dissolution is observed on the time scales of AIMD simulations.

## Acknowledgements

This research is supported by the U.S. Department of Energy, Office of Basic Energy Sciences under Grant No. DE-FG02-08ER46493.

## References

[1]    M. Gomina, P. Fourvel, M.H. Rouillon, High temperature mechanical behaviour of an uncoated SiC-SiC composite material, J. Mater. Sci. 26 (1991) 1891–1898.

[2]    S. Delpech, C. Cabet, C. Slim, G.S. Picard, Molten fluorides for nuclear applications, Mater. Today. 13 (2010) 34–41.

[3]    Y. Katoh, L.L. Snead, I. Szlufarska, W.J. Weber, Radiation effects in SiC for nuclear structural applications, Curr. Opin. Solid State Mater. Sci. 16 (2012) 143-152.

[4]    C. Forsberg, P.F. Peterson, H. Zhao, High-Temperature Liquid-Fluoride-Salt Closed-Brayton-Cycle Solar Power Towers, J. Sol. Energy Eng. 129 (2007) 141–146.

[5]    X. He, J. Song, J. Tan, B. Zhang, H. Xia, Z. He, X. Zhou, M. Zhao, X. Liu, L. Xu, S. Bai, SiC coating: An alternative for the protection of nuclear graphite from liquid fluoride salt, J. Nucl. Mater. 448 (2014) 1–3.

[6]    J. Roy, S. Chandra, S. Das, S. Maitra, Oxidation behaviour of silicon carbide-A review, Rev. Adv. Mater. Sci. 38 (2014) 29–39.

[7]    Y. Gu, J.X. Liu, Y. Wang, J.X. Xue, X.G. Wang, H. Zhang, F. Xu, G.J. Zhang, Corrosion behavior of TiC–SiC composite ceramics in molten FLiNaK salt, J. Eur. Ceram. Soc. 37 (2017) 2575–2582.

[8]    W. Xue, X. Yang, J. Qiu, H. Liu, B. Zhao, H. Xia, X. Zhou, P. Huai, H. Liu, J. Wang, Effects of $Cr^{3+}$ on the corrosion of SiC in LiF–NaF–KF molten salt, Corros. Sci. 114 (2017) 96–101.

[9]    Y. Liu, I. Szlufarska, Competition between strain and chemistry effects on adhesion of Si and SiC, Phys. Rev. B - Condens. Matter Mater. Phys. 79 (2009).

[10]   G. Cicero, A. Catellani, G. Galli, Atomic control of water interaction with biocompatible surfaces: The case of SiC(001), Phys. Rev. Lett. 93 (2004) 16102–1.

[11]   U. Cohen, R.A. Huggins, Silicon Epitaxial Growth by Electrodeposition from Molten Fluorides, J. Electrochem. Soc. (1971) 381–383.

[12]   A.L. Bieber, L. Massot, M. Gibilaro, L. Cassayre, P. Taxil, P. Chamelot, Silicon electrodeposition in molten fluorides, Electrochim. Acta. 62 (2012) 282–289.

[13]   G.M. Rao, D. Elwell, R.S. Feigelson, Electrocoating of silicon and its dependence on the time of electrolysis, Surf. Technol. 13 (1981) 331–337.

[14]   Y. Dong, T. Slade, M.J. Stolt, L. Li, S.N. Girard, L. Mai, S. Jin, Low-Temperature Molten-Salt Production of Silicon Nanowires by the Electrochemical Reduction of $CaSiO_3$, Angew. Chemie-Int. Ed. 56 (2017) 14453–14457.

[15]   A.R. Kamali, D.J. Fray, Molten salt corrosion of graphite as a possible way to make carbon nanostructures, Carbon. 56 (2013) 121–131.

[16]   A. Rezaei, A.R. Kamali, Green production of carbon nanomaterials in molten salts, mechanisms and applications, Diam. Relat. Mater. 83 (2018) 146-161.

[17]   S. News, S. News, Packmol: A Package for Building Initial Con gurations for Molecular Dynamics Simulations, J. Comput. Chem. (2009) 1–14.

[18]   A. Bengtson, H.O. Nam, S. Saha, R. Sakidja, D. Morgan, First-principles molecular dynamics modeling of the LiCl-KCl molten salt system, Comput. Mater. Sci. 83 (2014) 362–370.

[19]   H.O. Nam, A. Bengtson, K. Vörtler, S. Saha, R. Sakidja, D. Morgan, First-principles molecular dynamics modeling of the molten fluoride salt with Cr solute, J. Nucl. Mater. 449 (2014) 148–157.

[20]   M.P. Tosi, F.G. Fumi, Ionic sizes and born repulsive parameters in the NaCl-type alkali halides-II, J. Phys. Chem. Solids. 25 (1964) 45–52.

[21]  S. Plimpton, Fast parallel algorithms for short-range molecular dynamics, J. Comput. Phys. 117 (1995) 1–19.

[22]  G. Kresse, J. Furthmüller, Efficient iterative schemes for ab initio total-energy calculations using a plane-wave basis set, Phys. Rev. B - Condens. Matter Mater. Phys. 54 (1996) 11169–11186.

[23]  F. Vaslow, A.H. Narten, Diffraction pattern and structure of molten $BeF_2$-LiF solutions, J. Chem. Phys. 58 (1973) 5017.

[24]  G.J. Janz, Thermodynamic and Transport Properties for Molten Salts: Correlation Equations for Critically Evaluated Density, Surface Tension, Electrical Conductance, and Viscosity Data, New York, United States, 1988.

[25]  V. V. Ignat'ev, A. V. Merzlyakov, V.G. Subbotin, A. V. Panov, Y. V. Golovatov, Experimental investigation of the physical properties of salt melts containing sodium and lithium fluorides and beryllium difluoride, At. Energy. 101 (2006) 822–829.

[26]  D. Joubert, From ultrasoft pseudopotentials to the projector augmented-wave method, Phys. Rev. B-Condens. Matter Mater. Phys. 59 (1999) 1758–1775.

[27]  J.P. Perdew, K. Burke, M. Ernzerhof, Generalized gradient approximation made simple, Phys. Rev. Lett. 77 (1996) 3865–3868.

[28]  J. Klimeš, D.R. Bowler, A. Michaelides, Van der Waals density functionals applied to solids, Phys. Rev. B. 83 (2011) 195131.

[29]  J. Xi, B. Liu, Y. Zhang, W.J. Weber, Ab initio study of point defects near stacking faults in 3C-SiC, Comput. Mater. Sci. 123 (2016) 131–138.

[30]  H.O. Nam, D. Morgan, Redox condition in molten salts and solute behavior: A first-principles molecular dynamics study, J. Nucl. Mater. 465 (2015) 224–235.

[31]  J. Xi, B. Liu, Y. Zhang, W.J. Weber, Ab initio study of point defects near stacking faults in 3C-SiC, Comput. Mater. Sci. 123 (2016) 131–138.

[32]  M. Chase, NIST-JANAF Thermochemical Tables, 4th Edition, J. Phys. Chem. Ref. Data, Monogr. 9. (1998) 1952.

[33]  J. Xi, H. Xu, Y. Zhang, W.J. Weber, Strain effects on oxygen vacancy energetics in $KTaO_3$, Phys. Chem. Chem. Phys. 19 (2017) 6264–6273.

[34]  K. Reuter, M. Scheffler, Composition, structure, and stability of $RuO_2(110)$ as a function of oxygen pressure, Phys. Rev. B. 65 (2001) 035406.

[35]  K.S. B. Kelleher, K. Dolan, M. Anderson, Observed redox potential range of Li2BeF4 using a dynamic reference electrode, Nucl. Technol. 195 (2016) 239–252.

[36]  D. Olander, Redox condition in molten fluoride salts: Definition and control, J. Nucl. Mater. 300 (2002) 270–272.

[37]  S.N. Filimonov, Ab initio calculations of absolute surface energies of clean and hydrogen covered 3C-SiC(001), (110) and (111) surfaces, Mater. Sci. Forum. 821-823 (2015) 363-366.

[38]  M. Sabisch, P. Krüger, A. Mazur, M. Rohlfing, J. Pollmann, First-principles calculations of β-SiC(001) surfaces, Phys. Rev. B-Condensed Matter Mater. Phys. 53 (1996) 13121–13132.

[39]  V.L. Deringer, A.L. Tchougréeff, R. Dronskowski, Crystal orbital Hamilton population (COHP) analysis as projected from plane-wave basis sets, J. Phys. Chem. A. 115 (2011) 5461–5466.

[40]  X.D. Huang, J.K.O. Sin, P.T. Lai, Fluorinated $SrTiO_3$ as charge-trapping layer for nonvolatile memory applications, IEEE Trans. Electron Devices. 58 (2011) 4235–4240.

[41] S. Tinck, E.C. Neyts, A. Bogaerts, Fluorine-silicon surface reactions during cryogenic and near room temperature etching, J. Phys. Chem. C. 118 (2014) 30315–30324.

[42] X.D. Huang, R.P. Shi, P.T. Lai, Charge-trapping characteristics of fluorinated thin $ZrO_2$ film for nonvolatile memory applications, Appl. Phys. Lett. 104 (2014) 162905.

[43] M.M. Disko, J.C.H. Spence, O.F. Sankey, D. Saldin, Electron-energy-loss near-edge structure of $Be_2C$, Phys. Rev. B. 33 (1986) 5642–5651.

[44] J.T. Stock, M.V. Orna, Electrochemistry, past and present. Columbus, OH: American Chemical Society, 1989.

[45] M.S. Sohal, M. a Ebner, P. Sabharwall, P. Sharpe, Engineering database of liquid salt thermophysical and thermochemical properties, Idaho Natl. Lab. Idaho Falls CrossRef. (2010) 1–70.

[46] C. Laurence, J. Graton, J.F. Gal, An overview of Lewis basicity and affinity scales, J. Chem. Educ. 88 (2011) 1651–1657.

[47] C. Laurence, J.F. Gal, Lewis Basicity and Affinity Scales: Data and Measurement, 2009.

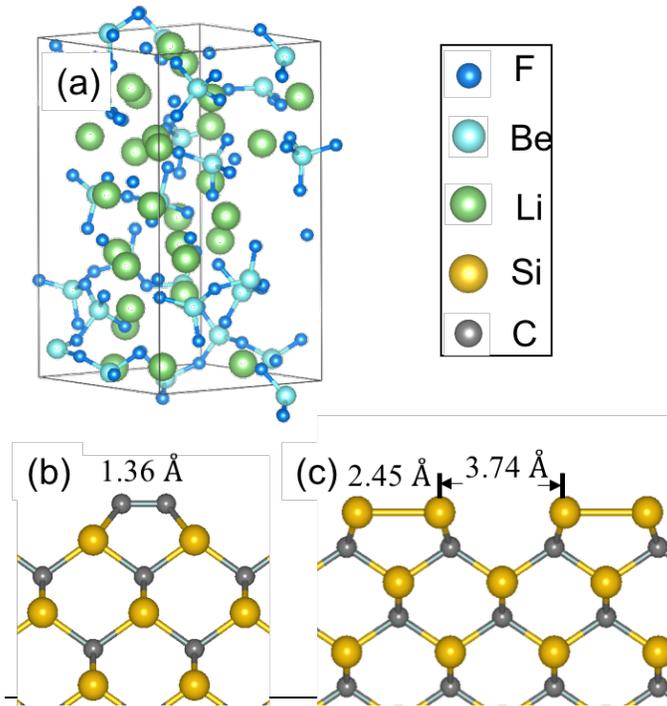

*Fig. 1. (a) Configuration of fully relaxed fluoride molten salt (2LiF-BeF$_2$). (b) and (c) SiC(001) surfaces in the p(2 × 1) reconstructions for C- and Si-termination, respectively. Dimer length and inter-dimer distances for C-terminated SiC(001) are 1.36 Å and 4.83 Å, respectively. Corresponding values for Si-terminated SiC(001) are 2.45 Å and 3.74 Å, respectively.*

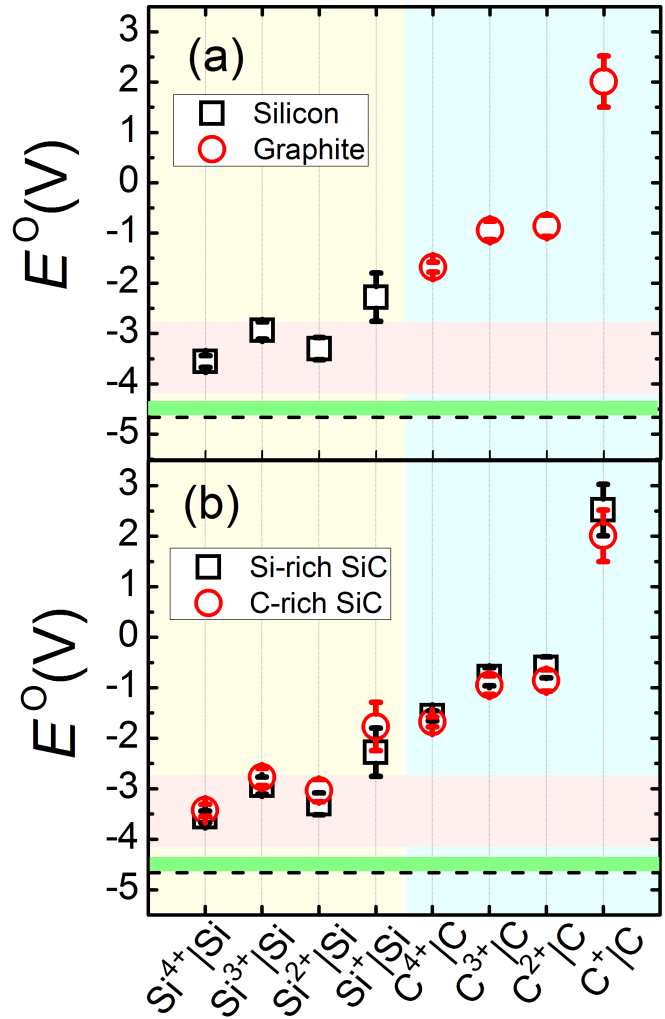

Fig. 2. A series of relevant potentials for FLiBe salts, with a reference potential set by the reaction of $F_2|F^-$. The calculated standard potential, $E^O$, for (a) silicon, graphite, and (b) SiC dissolution in FLiBe molten salt at 1000 K. The yellow region (left half) is for the dissolution reactions of Si species, the blue region (right half) for C species. The potentials for Si species in silicon are the same as those in SiC under Si-rich conditions, and the potentials for C species in graphite are the same as those in SiC under C-rich conditions. The red shaded horizontal rectangle corresponds to experimentally measured standard potential for FLiBe molten salt at 873 K [35]. The dashed line is the calculated standard potential, about -4.61 V (with standard deviation of 0.28 V), for metallic Be at 1000 K in FLiBe molten salt. The green horizontal rectangle is the calculated potential, from -4.61 V to -4.32 V (-4.58 V) (with standard deviation of 0.28 V) in the C-rich (Si-rich) conditions, for the reaction of $Be^0_{salt}|Be^{2+}$, within which solid $Be_2C$ could be formed.

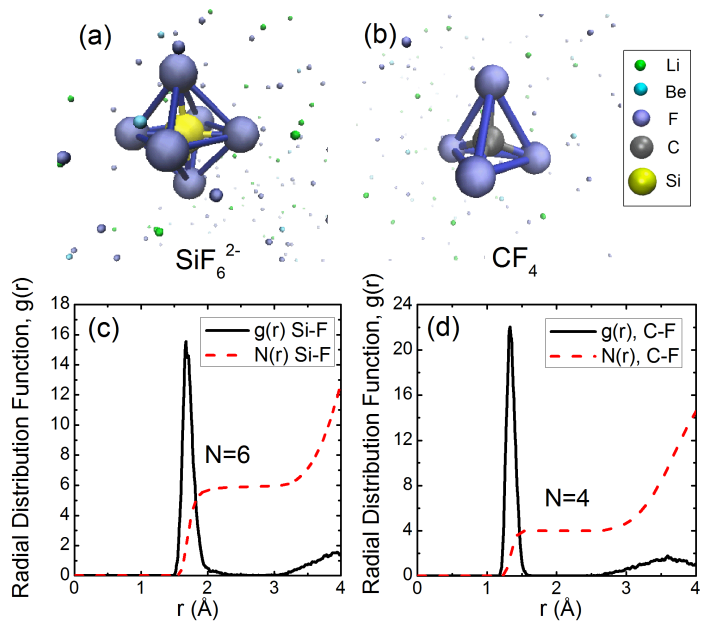

*Fig. 3. Snapshots of stable structure for (a) $Si^{4+}$ and (b) $C^{4+}$ solute ions in FLiBe salt. The corresponding radial distribution functions (RDF, g(r)) for (c) $Si^{4+}$ and (d) $C^{4+}$. Solid lines denote the RDF of solutes, dashed line represent the number of neighbors, N(r), between solute and F ions within a given cutoff distance r.*

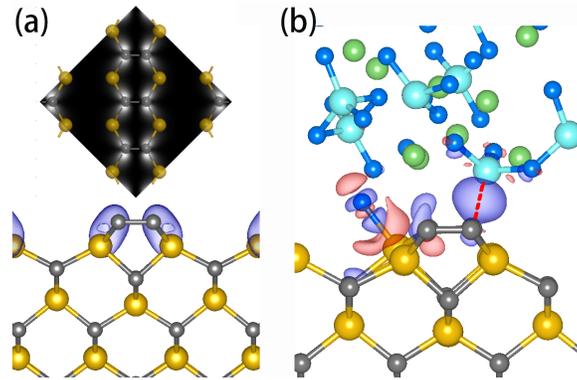

*Fig. 4. (a) Top and side views of the partial charge density distribution for C-terminated SiC(001) occupied surface states with energies 1.5 eV below the Fermi level. (b) The calculated charge density difference of C-terminated SiC(001)/salt. Red and blue regions depict the isocharge surface (at a value of $4.2 \times 10^{-2}$ e/Å³) of electron accumulation (positive) and depletion (negative). The colored atoms are depicted as in Fig. 1.*

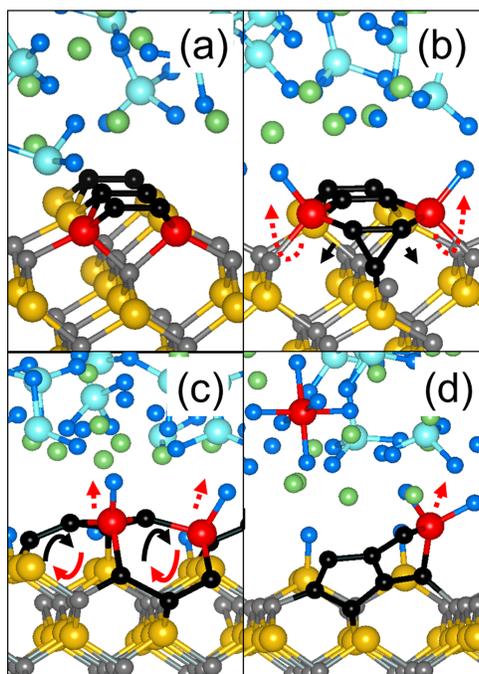

*Fig. 5. Snapshots of atomic configuration for the solid-salt interaction process on C-terminated SiC(001) surface at 2000 K. (a) and (b) the starting and intermediate configurations for the reaction of salt-surface, (c) and (d) the final configurations for the reaction of salt-surface with additional four and eight F atoms, respectively. The colored atoms are depicted as in Fig. 1. The black circles are the disordered carbon atoms, the red ones are the disordered silicon atoms.*

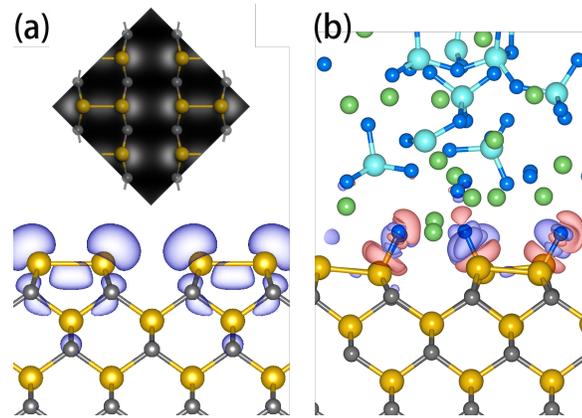

*Fig. 6. (a) Top and side views of the partial charge density distribution for Si-terminated SiC(001) occupied surface states with energies 1.5 eV below the Fermi level. (b) The calculated charge density difference of Si-terminated SiC(001)/salt. Red and blue regions in (a) and (b) depict the isocharge surface (at a value of $5.8 \times 10^{-2}$ $e/Å^3$) of electron accumulation (positive) and depletion (negative). Atoms are color-coded using the same scheme as in Fig. 1.*

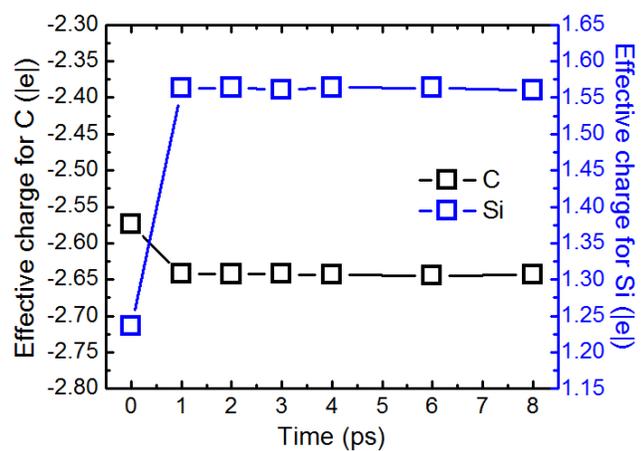

*Fig. 7. Time evolution of average effective charge for the first Si layer and the second C layer.*

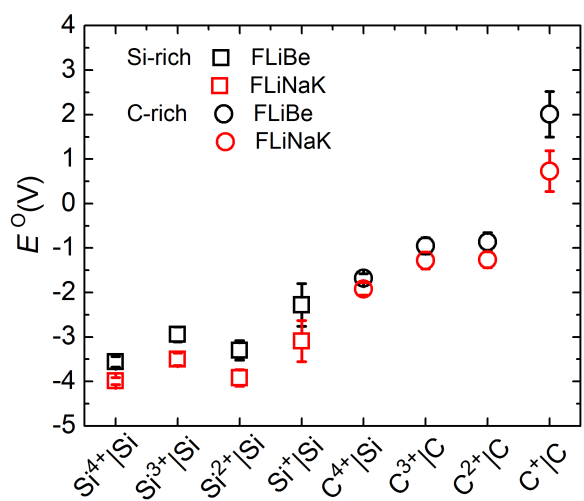

*Fig. 8. Comparison of the standard potentials for SiC dissolution at 1000 K in FLiBe and FLiNaK salts in Si- and C-rich conditions. The potentials for Si species in silicon are the same as those in SiC under Si-rich condition, and the potentials for C species in graphite are the same as those in SiC under C-rich condition.*